\documentstyle[12pt]{article}
\newfont{\sbfont}{cmssbx10 scaled\magstep1}
\def\QTR#1#2{{\csname#1\endcsname #2}}

\begin{document}
\vskip 1.8cm
\centerline{\bf USING FUNCTIONAL METHODS TO COMPUTE}
\vskip 0.6cm
\centerline{\bf QUANTUM EFFECTS IN THE LIOUVILLE MODEL\footnote{Based on
talks given at the
NATO Advanced Research Workshop, 7-12 January 1991, University of Miami.
Published in the proceedings,
{\it Quantum Field Theory, Statistical Mechanics, Quantum
Groups, and Topology}, T. Curtright, L. Mezincescu, and R. Nepomechie,
Ed's., World Scientific Publishing Co., 1992, pp 333-345.
We have been asked to make this paper more readily accessible,
so here it is.}}
\vskip 1.8cm
\centerline{\bf T. L. Curtright\footnote{\sf curtright@phyvax.ir.miami.edu}}
\vskip0.6cm
\centerline{Department of Physics, University of Miami}
\centerline{P. O. Box 248046, Coral Gables, FL 33124 USA}
\vskip0.6cm
\centerline{\bf G. I. Ghandour\footnote{\sf ghassan@kuc01.kuniv.edu.kw}}
\vskip0.6cm
\centerline{Faculty of Science, Physics Department, Kuwait University}
\centerline{P. O. Box 5969/13060, Safat, KUWAIT}

\vskip1.8cm

\centerline{\sf ABSTRACT}

We use time-independent canonical transformation methods to discuss the
energy eigenfunctions for the simple linear potential, pedagogically setting
the stage for some field theory calculations to follow. We then discuss the
Schr\"odinger wave-functional method of calculating correlation functions
for Liouville field theory. We compare this approach to earlier treatments,
in particular we check against known weak-coupling results for the Liouville
field defined on a cylinder. Finally, we further set the stage for future
Liouville calculations on curved two-manifolds and briefly discuss simple
quantum mechanical systems with time-dependent Hamiltonians.

\vfill

\newpage
\noindent QUANTUM MECHANICAL FOREPLAY

Consider a particle moving in a one-dimensional linear potential as
described by the Hamiltonian
\begin{equation}
H=p^2+q,
\end{equation}
where $q$ and $p$ are canonically conjugate position and momentum variables.
As every well-educated physicist knows, in the coordinate representation the
energy eigenfunctions for this problem are Airy functions, easily found by
solving the eigenvalue problem in the momentum representation and then
executing a Fourier transform. (cf. \S 24 in \cite{LandauLifschitz1977}.)

\ Nevertheless, let us construct the energy eigenfunctions from scratch
using the machinery of canonical transformations. Since the energy spectrum
is obviously continuous and unbounded either above or below, $-\infty
<E<+\infty $, it is clearly a good idea to map the above Hamiltonian onto a
free particle Hamiltonian with the same properties. So we take the free
Hamiltonian to be just the momentum itself.
\begin{equation}
H_{free}=P.
\end{equation}
Now, what is a generating function which produces a canonical transformation
from the ``interacting'' variables $(q,p)$ to the ``free'' variables $(Q,P)$
so that $H=H_{free}$? A simple answer is
\begin{equation}
F(q,Q)=-\frac 13Q^3-qQ.
\end{equation}
We may check this statement at the classical level as follows.
\begin{equation}
p_{classical}={\frac \partial {\partial q}}F=-Q,~~~~~~~~P_{classical}=-{%
\frac \partial {\partial Q}}F=(Q^2+q).
\end{equation}
Hence $p_{classical}^2+q=P_{classical}$ as anticipated.

The significance of the above for the quantum mechanical problem was first
appreciated in a general setting by Dirac, in his famous 1932 article \cite
{Dirac1932} which motivated Feynman's development of the path integral,
although Dirac was not inclined to provide any details for such simple
examples as that of the case at hand.%
\footnote{Some other simple examples have been explicitly discussed by
G. Ghandour, Phys. Rev. \QTR{bf}{D35} (1987) 1289. Also see
T.L. Curtright, \QTR{it}{Differential Geometric Methods in Theoretical Physics:
Physics and Geometry}, L.-L. Chau and W. Nahm, ed's.,
Plenum Press, 1990, pp 279-289.} If we follow Dirac's advice and
exponentiate the above generating function we obtain a transformation
functional which produces energy eigenfunctions in coordinate space for the
linear potential, $\Phi _E(q)$, from those for the free particle, $\Psi
_E(Q) $. Thus
\begin{equation}
\Phi _E(q)=\int_{-\infty }^{+\infty }dQ~e^{iF(q,Q)}~\Psi _E(Q).
\end{equation}
The free particle eigenfunctions are given here by $\Psi _E(Q)={\frac 1{2\pi
}}~e^{iEQ}$.

We easily check that the exponentiated generating function provides an exact
solution of the coordinate space partial differential equation
\begin{equation}
\label{HexpiF}H~e^{iF}=-H_{free}~e^{iF},
\end{equation}
where
\begin{equation}
H=-{\frac{\partial ^2}{\partial q^2}}+q,~~~~~~~~H_{free}=-i{\frac \partial
{\partial Q}}.
\end{equation}
Then if we integrate by parts and discard terms at $Q=\pm \infty $, we
verify the above relation between $\Phi _E(q)$ and $\Psi _E(Q)$.
$$
H~\Phi _E(q)=\int_{-\infty }^{+\infty }dQ~\left( H~e^{iF(q,Q)}~\right) \Psi
_E(Q)=\int_{-\infty }^{+\infty }dQ~\left( -H_{free}~e^{iF(q,Q)}\right) ~\Psi
_E(Q)
$$
\begin{equation}
=\int_{-\infty }^{+\infty }dQ~e^{iF(q,Q)}~H_{free}~\Psi _E(Q)=E~\Phi _E(q).
\end{equation}
Of course, we recognize in all this the well-known integral representation
for the Airy function.
\begin{equation}
Ai(x)={\frac 1{2\pi }}~\int_{-\infty }^{+\infty }dz~e^{-i(z^3/3+zx)}={\
\frac 1\pi }~\int_0^{+\infty }dz~\cos (z^3/3+zx).
\end{equation}
That is, $\Phi _E(q)=Ai(q-E)$. This is precisely the representation for the
energy eigenfunctions which follows by the textbook momentum-space/Fourier
transform approach mentioned earlier.%
\footnote{Interestingly, as a well-informed reader
will further recognize, $F(q,Q)$ is the
``canonical unfolding'' of the elementary catastrophe germ $G = Q^3$.
(For example, see R. Gilmore,
\QTR{it}{Catastrophe Theory for Scientists and Engineers},
John Wiley \& Sons, 1981.)
We note in passing that
the canonical unfoldings of all seven elementary catastrophes
(with co-dimension $\leq 4$) provide generating functions for canonical
transformations from interacting systems involving linear potentials
to free systems, in complete parallel with the development which
we have given here for $F(q,Q)$.}

This integral expression for the linear potential energy eigenfunctions, as
provided by the generating function of a canonical transformation, is quite
useful. It allows us to compute correlation functions involving the
interacting variables $q$ and $p$ through a reduction to correlation
functions involving the free variables $Q$ and $P$. As an illustration
consider the propagator
\begin{equation}
\Delta (q_1,q_2;t)=\int_{-\infty }^{+\infty }dE~e^{-iEt}~\Phi _E(q_1)~\Phi
_E^{*}(q_2).
\end{equation}
For the linear potential problem, a closed-form result for $\Delta $ follows
immediately from the above integral representation. Although the generating
function is cubic in $Q$, it is still only necessary to evaluate Gaussian
integrals as evident in the following steps.
$$
\Delta (q_1,q_2;t)={\frac 1{4\pi ^2}}~\int_{-\infty }^{+\infty
}dE~\int_{-\infty }^{+\infty }dQ_1~\int_{-\infty }^{+\infty
}dQ_2~e^{-iE(t-Q_1+Q_2)}~e^{-i(Q_1^3/3+q_1Q_1)}~e^{i(Q_2^3/3+q_2Q_2)}
$$
$$
={\frac 1{2\pi }}~\int_{-\infty }^{+\infty }dQ_1~\int_{-\infty }^{+\infty
}dQ_2~\delta (t-Q_1+Q_2)~e^{-i(Q_1^3/3+q_1Q_1)}~e^{i(Q_2^3/3+q_2Q_2)}
$$
\begin{equation}
={\frac 1{2\pi }}~\int_{-\infty }^{+\infty
}dQ_1~e^{-i(Q_1^3/3+q_1Q_1)}~e^{i((Q_1-t)^3/3+q_2(Q_1-t))}.
\end{equation}
So the terms cubic in $Q_1$ cancel in the exponent, which leaves just a
Gaussian in $Q_1$ and hence the result. 
\begin{equation}
\Delta (q_1,q_2;t)={\frac 1{2\pi }}~\sqrt{{\frac \pi {it}}}~\exp {\left( {\
\frac{i(q_1-q_2)^2}{4t}}-{\frac{i(q_1+q_2)t}2}-{\frac{i~t^3}{12}}\right) }.
\end{equation}
The reader is encouraged to compare this method of computing the propagator
with the more familiar path integral method \cite{FeynmanHibbs1965}.

There are two special features of the linear potential problem which are
{\it not} generally true \cite{Ghandour1987}. Firstly, the expression for
the classical generating function $F$ works to solve exactly the quantum
mechanical linear potential problem. In general, this is not the case as
there are usually quantum corrections (i.e. terms O($\hbar $)) which must be
added to $F$. Secondly, both the normalization and overall phase of the
energy eigenfunction are unchanged by the canonical transformation for the
linear potential. In general, this is again not the case as it is usually
necessary to introduce an energy dependent normalization and phase factor $%
N(E)$, so
\begin{equation}
\Phi _E(q)=N(E)~\int_{-\infty }^{+\infty }dQ~e^{iF(q,Q)}~\Psi _E(Q).
\end{equation}
We shall next discuss an example, Liouville field theory, where both of
these more general features arise. Fortunately, the resulting modifications
are so simple in the Liouville case that most if not all of the essential
information is still contained in the classical generating function.

\vskip 0.6cm \noindent{FUNCTIONAL APPROACH TO LIOUVILLE THEORY}

Liouville field theory consists of a scalar field $\phi $ in exponential
self-interaction. In this section, we will consider Liouville theory on a
two-dimensional spacetime cylinder, $(0\le \sigma \le 2\pi ,-\infty <\tau
<+\infty )$, subject to periodic boundary conditions in $\sigma $. As
Liouville himself demonstrated for the classical theory, this interacting
field is canonically equivalent to a free pseudoscalar field $\psi $. The
equivalence for fixed time, $\tau $, is most readily apparent from the
existence of a generating functional
\begin{equation}
F[\phi ,\psi ]=\int_0^{2\pi }d\sigma ~\left( \phi ~\partial _\sigma \psi -{\
\frac{2m}{g^2}}~e^{g\phi }\sinh (g\psi )\right) .
\end{equation}
For the classical theory, canonical momenta are obtained from $F$ in the
usual way through functional derivation.
\begin{equation}
\pi _{\phi ,classical}={\frac{\delta F}{\delta \phi }},\qquad \pi _{\psi
,classical}=-{\frac{\delta F}{\delta \psi }}.
\end{equation}
Evaluation of these derivatives leads to a pair of first-order equations,
the so-called B\"acklund transformations for the classical theory.
$$
\partial _\tau \phi =\pi _{\phi ,classical}=\partial _\sigma \psi -{\frac{2m}%
g}~e^{g\phi }\sinh (g\psi ),
$$
\begin{equation}
\label{Backlund}\partial _\tau \psi =\pi _{\psi ,classical}=\partial _\sigma
\phi +{\frac{2m}g}~e^{g\phi }\cosh (g\psi ).
\end{equation}
Integrability conditions for these equations follow from alternately
differentiating with respect to $\sigma $ and $\tau $. Consistency requires
that the classical $\phi $ and $\psi $ fields must obey the Liouville and
free wave equations.
\begin{equation}
\label{WaveEqns}({\partial _\tau }^2-{\partial _\sigma }^2)\,\phi +{\frac{%
4m^2}g}~e^{2g\phi }=0,\qquad ({\partial _\tau }^2-{\partial _\sigma }%
^2)\,\psi =0.
\end{equation}

For the classical Liouville field theory, the B\"acklund equations may be
integrated to obtain all solutions for the interacting field $\phi$ in terms
of the free field $\psi$. Hence, all interacting field functionals $%
g[\phi,\pi_{\phi}]$ may be expressed in terms of free field functionals, $%
G[\psi,\pi_{\psi}]$, and the classical theory is completely solved. However,
there remains a long way to go to arrive at a quantum Liouville theory, even
with a complete set of classical solutions.

There are at least three major routes one can take to arrive at a quantized
Liouville theory. These involve either operator methods \cite
{CurtrightThorn1982}, Schr\"odinger functional techniques \cite
{CurtrightGhandour1984}, or path integrals \cite{GoulianLi1991}. Among these
three major routes, there are also several minor variations, and while all
these approaches may indeed be equivalent in principle, they are not
necessarily equivalent in practice.

For example, one operator approach to quantize the theory is to convert the
classical relations between $\phi $ and $\psi $ into well-defined operator
expressions in such a way that the locality and conformal transformation
properties expected of the expressions do indeed hold. This is a laborious
procedure, but it has been carried out for many of the classical relations
between Liouville and free fields \cite{BraatenCurtrightThorn1983}.
Unfortunately, even when valid operator relations have been obtained, there
still remains the difficult task of using those operator results to evaluate
correlation functions.

Here we shall follow the canonical functional methods pedagogically
discussed in the previous section for the linear potential. We will use the
generating functional $F[\phi ,\psi ]$ within the Schr\"odinger equal-time
functional formalism to construct energy eigenfunctionals $\Phi _E[\phi ]$
for the quantum Liouville theory from well-known free field energy
eigenfunctionals $\Psi $$_E[\psi ]$. We then approach the task of evaluating
correlation functions in a fashion completely analogous to the evaluation of
the propagator for the linear potential. As in that simple case, the end
result for the Liouville theory, at least in principle, is to reduce the
problem to the evaluation of free field functional integrals. Indeed, in the
weak-coupling limit, we immediately obtain a series of Gaussian functional
integrals which evaluates to reproduce results obtained using operator
methods \cite{BraatenCurtrightGhandourThorn1984}. Again, as in the case of
the propagator for the linear potential, this functional approach to
correlation functions should be compared to that using path integrals \cite
{GoulianLi1991}. Unfortunately, here we will be forced to leave a detailed
comparison with path integral results as an exercise for the interested
student. Suffice it to say that the functional methods seem to be easier to
implement than operator methods, and they may provide a useful bridge
between operator methods and path integral techniques.

Let us return to the B\"acklund transformations, Eqn.(\ref{Backlund}). These
first-order functional derivative relations are most conveniently written
for the quantum Liouville theory as
\begin{equation}
D_{\pm }(\sigma )~e^{iF}=0,
\end{equation}
where
\begin{equation}
D_{\pm }(\sigma )\equiv -i{\left( {\frac \delta {\delta \phi (\sigma )}}~\pm
{}~{\frac \delta {\delta \psi (\sigma )}}\right) }~-\left( \partial _\sigma
\psi (\sigma )~\mp ~\partial _\sigma \phi (\sigma )\right) +{\frac{2m}g}%
{}~e^{g\phi (\sigma )\pm g\psi (\sigma )}\;.
\end{equation}
Considering two such functional derivatives, in the form $D_{\pm }(\sigma
_1)D_{\pm }(\sigma _2)e^{iF}$ as $\sigma _1\rightarrow \sigma _2$, leads to
the conclusion that the Liouville and free field Hamiltonians have
equivalent effects on $e^{iF}$, just as in the case of the linear potential,
Eqn.(\ref{HexpiF}). Consequently, the Liouville energy eigenfunctionals are
functional transforms of the free field eigenfunctionals.%
\footnote{
N.B.  The field integrations in this result, and in the expectation
of $\exp(\alpha g \phi)$, are over all field configurations
at fixed time.  That is, $\int d\phi$ and $\int d\psi$ here,
and elsewhere in this paper, are {\sl not} path integrals,
but rather Schr\"odinger functional integrals.}
\begin{equation}
\label{FunctionalTransform}\Phi _E[\phi ]=N(E)~\int d\psi ~e^{iF[\phi ,\psi
]}~\Psi _E[\psi ].
\end{equation}
As in the previous quantum mechanics example, the classical{\em \ form} for
the generating functional serves to provide an exact transformation between
interacting and free theories. However, there is one important difference
between the simple quantum mechanics example and the Liouville theory. The
parameters in $F$ are renormalized, as explained below.

In any case, correlation functions are now given in terms of functional
integrals. For example, the expectation of an exponentiated Liouville field
is
$$
\langle E_1|e^{\alpha g\phi }|E_2\rangle =\int d\phi ~\Phi _{E_1}^{*}[\phi
]~e^{\alpha g\phi }~\Phi _{E_2}[\phi ]
$$
\begin{equation}
=N(E_1)^{*}N(E_2)~\int d\psi _1~\int d\psi _2~\Psi _{E_1}^{*}[\psi
_1]~O[\psi _1,\psi _2]~\Psi _{E_2}[\psi _2]
\end{equation}
where the effective operator on the space of $\psi $ functionals is
\begin{equation}
O[\psi _1,\psi _2]=\int d\phi ~\exp \left( -iF^{*}[\phi ,\psi _1]+iF[\phi
,\psi _2]+\alpha g\phi \right) .
\end{equation}
In principle, this should be equivalent to the operator results of Braaten
et al. \cite{BraatenCurtrightThorn1983}, but in practice, we believe that
this functional form for the expectation value can sometimes lead to more
immediate results. For example, perturbation theory in the non-zero mode
effects for expectations between low energy states is immediately developed
from the functional expression.

The traditional separation of the fields into zero ($q$ and $Q)$ and
non-zero ($\hat \phi $ and $\hat \psi $) modes at fixed $\tau $ is given by
\cite{BraatenCurtrightThorn1983}
\begin{equation}
\phi (\sigma )=q+\hat \phi (\sigma ),\qquad \hat \phi (\sigma )={\frac i{
\sqrt{4\pi }}}~\sum_{n\ne 0}{\frac 1n}~(a_ne^{-in\sigma }+b_ne^{in\sigma }),
\end{equation}
\begin{equation}
\psi (\sigma )=Q+\hat \psi (\sigma ),\qquad \hat \psi (\sigma )={\frac i{
\sqrt{4\pi }}}~\sum_{n\ne 0}{\frac 1n}~(A_ne^{-in\sigma }+B_ne^{in\sigma }).
\end{equation}
However, one should keep in mind that the expansion coefficients in the
functional formalism used here are{\em \ not }operator valued.

For the free field, we may explicitly display the time dependence of the
modes to obtain right-movers
\begin{equation}
\hat \psi _{\Rightarrow }(\tau ,\sigma )={\frac i{\sqrt{4\pi }}}~\sum_{n\ne
0}{\frac 1n\,}~B_ne^{-in(\tau -\sigma )},
\end{equation}
and left-movers
\begin{equation}
\hat \psi _{\Leftarrow }(\tau ,\sigma )={\frac i{\sqrt{4\pi }}}~\sum_{n\ne 0}%
{\frac 1n}~\,A_ne^{-in(\tau +\sigma )}.
\end{equation}
For $\hat \phi $, the time-dependence is not so simple. Rather, it is
convenient to combine the left- and right-moving modes for fixed time using
projection operators. We first define these projections for the free field.
\begin{equation}
\hat \psi (\sigma )=\sum_{n\ne 0}\psi _n\;e^{in\sigma }=\hat \psi
_{+}(\sigma )+\hat \psi _{-}(\sigma ),\;\;\;\;\;\;\;\;\;\hat \psi _{\pm
}(\sigma )=P_{\pm }\;\hat \psi (\sigma )=\sum_{n>0}\psi _{\pm n}\;e^{\pm
in\sigma },
\end{equation}
\begin{equation}
P_{\pm }\equiv \frac 12\left( 1\mp i\frac{\partial _\sigma }{\left| \partial
_\sigma \right| }\right) ,\;\;\;\;\;\;\;\;\;\;\;\;\;\;|\partial _\sigma
|\equiv \sqrt{-{\frac{\partial ^2}{\partial \sigma ^2}}}.
\end{equation}
We then note that this separation of the modes is a well-defined procedure
even for the interacting field.
\begin{equation}
\hat \phi (\sigma )=\sum_{n\ne 0}\phi _{n\;}e^{in\sigma }=\hat \phi
_{+}(\sigma )+\hat \phi _{-}(\sigma ),\;\;\;\;\;\;\;\;\hat \phi _{\pm
}(\sigma )=P_{\pm }\;\hat \phi (\sigma )=\sum_{n>0}\phi _{\pm n}\;e^{\pm
in\sigma }.
\end{equation}
Using this separation into zero and non-zero modes, we may now parallel the
operator approach in Braaten et al. \cite{BraatenCurtrightGhandourThorn1984}
and perform a perturbative analysis of the effects of the non-zero modes for
low energy expectation values of exponentiated Liouville fields.

The non-zero mode free field vacuum functional is
\begin{equation}
\Psi _{vacuum}[\hat \psi ]=\exp \left( -{\frac 12}\int_0^{2\pi }d\sigma
{}~\hat \psi (\sigma )~|\partial _\sigma |~\hat \psi (\sigma )\right) ,
\end{equation}
where $|\partial _\sigma |~\psi (\sigma )=|\partial _\sigma |~\hat \psi
(\sigma )=\sum_{n>0}\left| n\right| (\psi _{+n}\;e^{+in\sigma }+
\psi _{-n}\;e^{-in\sigma })$.
{}From this we construct low energy free field wave functionals, $\Psi _k[\psi
]=\Psi _k[Q]\,\,\Psi _{vacuum}[\hat \psi ]$, with $E\equiv g^2k^2/4\pi $ and
$\Psi _k[Q]=N_k~\exp (i\,gkQ)$ , where it is understood that $g$ is small
and $k$ is of order unity. The functional transform (\ref
{FunctionalTransform}) then yields low energy Liouville wave functionals.
\begin{equation}
\Phi _k[\phi ]=N_k~\int dQ\;\exp (i\,gkQ)\;\int d\hat \psi ~e^{iF[\phi ,\psi
]}~\Psi _{vacuum}[\hat \psi ].
\end{equation}
We now split-off the non-zero mode contributions by writing
\begin{equation}
F[\phi ,\psi ]=\int_0^{2\pi }d\sigma ~\left( \phi ~\partial _\sigma \psi -{\
\frac{2m}{g^2}}~e^{gq}\sinh (gQ)\right) +F_{int}[\phi ,\psi ],
\end{equation}
where the interaction between zero and non-zero modes is contained in
\begin{equation}
\label{Fint}F_{int}=-{\frac m{g^2}}~e^{g(q+Q)}~\int_0^{2\pi }d\sigma ~\left(
e^{g(\hat \phi +\hat \psi )}-1\right) -{\frac m{g^2}}~e^{g(q-Q)}~\int_0^{2%
\pi }d\sigma ~\left( e^{g(\hat \phi -\hat \psi )}-1\right) .
\end{equation}
We shift $\hat \psi \to \hat \psi +\hat \phi _{+}-\hat \phi _{-}$, complete
the square, and eliminate the $\int d\sigma ~\hat \phi ~\partial _\sigma
\hat \psi $ term to obtain
$$
\Phi _k[\phi ]=N_k~\exp \left( -{\frac 12}\int_0^{2\pi }d\sigma ~\hat \phi
(\sigma )~|\partial _\sigma |~\hat \phi (\sigma )\right) \;\int dQ~\exp
\left( i\,gkQ\;-{\frac{4\pi mi}{g^2}}~e^{gq}\sinh (gQ)\right) \times
$$
\begin{equation}
{}~\times ~\int d\hat \psi ~\exp \left( iF_{int}[\phi ,\psi +\hat \phi
_{+}-\hat \phi _{-}]-{\ \frac 12}\int_0^{2\pi }d\sigma ~\hat \psi (\sigma
)~|\partial _\sigma |~\hat \psi (\sigma )\right) .
\end{equation}
After the shift, the interaction becomes%
$$
F_{int}[\phi ,\psi +\hat \phi _{+}-\hat \phi _{-}]=
$$
\begin{equation}
-\,{\frac m{g^2}}~e^{g(q+Q)}~\int_0^{2\pi }d\sigma ~\left( e^{g(2\hat \phi
_{+}+\hat \psi )}-1\right) -{\frac m{g^2}}~e^{g(q-Q)}~\int_0^{2\pi }d\sigma
{}~\left( e^{g(2\hat \phi _{-}-\hat \psi )}-1\right) .
\end{equation}
Now expand $\exp (iF_{int})$ in powers of $F_{int}$ to obtain a series of
Gaussian integrals
$$
\int d\hat \psi ~\exp \left( iF_{int}[\phi ,\psi +\hat \phi _{+}-\hat \phi
_{+}]-{\ \frac 12}\int_0^{2\pi }d\sigma ~\hat \psi (\sigma )~|\partial
_\sigma |~\hat \psi (\sigma )\right)
$$
\begin{equation}
={\sum_{n=0}^\infty \frac{i^n}{n!}}~\int d\hat \psi ~e^{-{\frac 12}%
\int_0^{2\pi }d\sigma ~\hat \psi (\sigma )~|\partial _\sigma |~\hat \psi
(\sigma )}~\left( F_{int}[\phi ,\psi +\hat \phi _{+}-\hat \phi _{-}]\right)
^n
\end{equation}
Thus $\Phi _k[\phi ]$ reduces to a series of $\hat \psi $ Gaussian integrals
with coefficients which depend on the zero modes, $q$ and $Q$. Subsequent
integration over $Q$ yields a $q$ and $\hat \phi $ dependent series.
\begin{equation}
\Phi _k[\phi ]=\sum_{n=0}^\infty \Phi _k^{(n)}[q,\hat \phi ]
\end{equation}
In this series, $\Phi ^{(n)}$ is obtained by keeping only the term involving
$(F_{int})^n$ in the previous expansion.

\ The lowest order non-trivial results for the expectation of $e^{\alpha
g\phi }$ are obtained by keeping terms up to and including $(F_{int})^{3}$
in this series. The evaluation of the resulting Gaussian integrals
is straightforward. To O($g^6$) we obtain

$$
\langle \Phi _{k_1}|e^{\alpha g\phi }|\Phi _{k_2}\rangle =\int d\phi ~\Phi
_{k_1}^{*}[\phi ]~e^{\alpha g\phi }~\Phi _{k_2}[\phi ]
$$
$$
=N_{k_1}\,N_{k_2}\,Z_{k_1k_2}(\alpha ,m,g)\,e^{\alpha ^2g^2\zeta _1(N)/2\pi
}\,\left( 1+8\left( \alpha -(k_1^2+k_2^2)\right) \left( \frac{g^2}{4\pi }%
\right) ^2\zeta _2\right.
$$
\begin{equation}
\label{<exp>}\left. +\left( \alpha (2-\alpha )(\frac 43+\frac 23\alpha
+\alpha ^2)+2\alpha (2-\alpha )(k_1^2+k_2^2)-(k_1^2-k_2^2)^2\right) \left(
\frac{g^2}{4\pi }\right) ^3\zeta _3+O(g^8)\right) ,
\end{equation}
where the zero-mode matrix element is defined by
\begin{equation}
\label{Z}Z_{k_1k_2}(\alpha ,m,g)=\frac{2e^{\pi (k_2-k_1)}}{g^3\Gamma (\alpha
)}\left( \frac{g^2}{2\pi m}\right) ^\alpha \ \left| \Gamma \left( \frac{%
\alpha +i(k_1+k_2)}2\right)
{}~\Gamma \left( \frac{\alpha
+i(k_1-k_2)}2\right) \right| ^2,
\end{equation}
and where $N_k$ is a normalization factor required by the zero-mode wave
function. The exact form of $N_k$ is not important for the present
discussion, but the possibility of such normalization factors is important,
as we shall see. The usual ultraviolet divergence in the vacuum expectation
value of an ``un-normal-ordered'' exponential is present here and is given
by
\begin{equation}
e^{\alpha ^2g^2\zeta _1(N)/2\pi }=\int d\hat \psi ~e^{\alpha g\hat \psi
}\,\mid \Psi _{vacuum}[\hat \psi ]\mid ^2
\end{equation}
where we have imposed a mode cutoff, $N$ , to write $\zeta
_1(N)=\sum_{n=1}^N\frac 1n$ .

Actually, to obtain (\ref{<exp>}), it is necessary to remove a similar
divergence due to the exponentials in the generating functional. To this end
we have replaced $e^{g\hat \phi }$ appearing in (\ref{Fint}) by $%
e^{-g^2\zeta _1(N)/2\pi }\,e^{g\hat \phi }$. We may think of this as a
renormalization of the ``mass'' appearing in $F_{int}$ . This is not an
unexpected renormalization. However, it is not the whole story for mass
renormalization, as is evident from the O($g^4$) terms in (\ref{<exp>}). The
$\alpha \zeta _2$ term on the RHS is a problem (the quantum version of the
equations of motion would fail) and must be eliminated. This may be achieved
by making an additional finite renormalization of the mass in $F$ , which
acquires an $\alpha $ dependence and cancels the $\alpha \zeta _2$ term
through its appearance in the zero-mode expression (\ref{Z}). The structure
of certain terms in the perturbation series suggests that this finite mass
renormalization is of the form $\frac 2{g^2}\,\sin (g^2/2)$, although we
have only checked this fully to O($g^4$). Nonetheless, this form is
supported to all orders by comparison with the operator results in Braaten
et al.\cite{BraatenCurtrightThorn1983}. In summary, the mass appearing in $F$
is renormalized according to
\begin{equation}
m\rightarrow e^{-g^2\zeta _1(N)/2\pi }\;\xi \;m\;,\;\;\;\;\;\;\;\;\;\;\xi
\equiv \left( \frac 2{g^2}\,\sin (g^2/2)\right) .
\end{equation}
Based on exact results within the operator formalism and for the effective
potential of the Liouville theory, we also anticipate another quantum
correction to the exponentials in $F$, in the form of a finite
renormalization of $g$.
\begin{equation}
g\rightarrow \frac g{1+\frac{g^2}{2\pi }}\;.
\end{equation}
However, we have not checked this correction within the functional
formalism. To do so to lowest order would require perturbative results
beyond O($g^6$).

Finally, other anomalous O($g^4$) non-zero mode contributions in (\ref{<exp>}%
) can be eliminated through changes in normalizations. The terms $%
(k_1^2+k_2^2)\left( \frac{g^2}{4\pi }\right) ^2\zeta _2$ may be removed by
changing the normalizations of the states $\Phi _{k_1}^{*}[\phi ]~$and $\Phi
_{k_2}[\phi ]~$. At this time, we have no firm conjectures about the form
for such normalization changes in higher orders.

After these renormalizations, the results in (\ref{<exp>}) agree to O($g^6$)
with those of Braaten et al. \cite{BraatenCurtrightGhandourThorn1984}
obtained through the use of operator techniques. Perhaps it will be possible
to extend these perturbative results to obtain correct closed-form
expressions to all orders in $g$ within the functional framework. Or,
perhaps it will be possible to numerically evaluate the functional
expressions to obtain non-perturbative results. We leave these as open
problems for the interested reader.

\vskip 0.6cm \noindent{LIOUVILLE THEORY ON CURVED SURFACES}

Perhaps functional methods are also useful when the $(\tau ,\sigma
)=(z^0,z^1)$ manifold is not intrinsically flat. The classical relations
between $\phi $ and $\psi $ have been discussed for this case by Preitschopf
and Thorn \cite{PreitschopfThorn1991}. Classically, these fields now satisfy
the covariant equations%
\footnote{Again, the $\frac 1g$ coefficients of $R$ and $\omega _\mu $
are expected to be modified by quantum effects to $\frac 1g+\frac g{2\pi}$.}
\begin{equation}
D^\mu D_\mu \phi ={\frac 1{2g}}~R-{\frac{4m^2}g}~e^{2g\phi },\ \ \ \ \ \
D^\mu (\partial _\mu \psi -{\frac 1g}~\omega _\mu )=0,
\end{equation}
where the connection and scalar curvature,
\begin{equation}
\omega _\mu =\eta _{ab}e_\mu ^a\epsilon ^{\nu \lambda }\partial _\nu
e_\lambda ^b,\ \ \ \ \ \ \ \ R=-2\epsilon ^{\mu \nu }\partial _\mu \omega
_\nu ,
\end{equation}
and the zweibein $e_\mu ^a$ are given functions that depend on $(z^0,z^1)$.

Canonical equivalence of the $\phi $ and $\psi $ fields in this curved
surface situation again follows from a generating functional. In this case,
as opposed to the cylinder, that generating functional depends explicitly on
$z^0$ since the zweibein and connection do.
\begin{equation}
\label{CurvedF}F[\phi ,\psi ;z^0]=\int dz^1\left( \phi \,\left( \partial
_1\psi -{\frac 1g}~\omega _1(z)\right) -\frac{2m}{g^2}\,e^{g\phi
}\,e_1^a(z)V_a(\psi )\right) \
\end{equation}
where the tangent space vector $V$ is given by $(V_0,V_1)=\left( \cosh
(g\psi ),\sinh (g\psi )\right) $. Therefore, to understand the quantum
theory in this situation, we require an extension of the previous formalism
to the case where the generating functional has explicit time dependence. In
principle, this extension is known, but in practice, this is still an open
problem for the Liouville quantum theory, which we intend to discuss further
elsewhere. Here it will suffice to return to the simple setting of quantum
mechanics to discuss the case of generating functionals with explicit time
dependence, again in the context of the linear potential.

\vskip 0.6cm \noindent TIME-DEPENDENT QUANTUM MECHANICS

It is not difficult to extend the above functional methods for solving
quantum mechanical models to the case of systems with explicitly
time-dependent generating functionals. In fact, the free particle is
probably the most familiar example of the method which involves just such an
extension. Let us briefly review this situation. As is well-known, the
general solution for the time-dependent wave function of a free particle
moving in one dimension can be written as $\Psi (q,t)=\int_{-\infty
}^{+\infty }dQ~{\frac{e^{i(q-Q)^2/2t}}{\sqrt{2\pi it}}}~\Psi _0(Q),$ where
the initial data is specified by $\Psi _0(Q)$. We may view this as a
canonical transformation, from the variable $q$, whose time development is
given by the Hamiltonian $H={\frac 12\,}p^2,$ to the variable $Q$, whose time
development is nil (i.e. $H_0\equiv 0$). As such, we may write the wave
function at time $t$ as $\Psi (q,t)=\int_{-\infty }^{+\infty
}dQ~e^{iF(q,Q;t)}~\Psi _0(Q).$ The quantum generator of the transformation
is $F(q,Q;t)={\frac{(q-Q)^2}{2t}}+{\frac i2}~\ln(2\pi it),$ and it provides a
solution of the evolution equation $\left( H-i{\frac \partial {\partial t}}%
\right) ~e^{iF(q,Q;t)}=H_0~e^{iF(q,Q;t)}\equiv 0,$ where the Hamiltonian $H$
is written in the coordinate basis, $H=-{\frac 12}~{\frac{\partial ^2}{%
\partial q^2}.}$ Note that the $\ln(2\pi it)$ term in $F$ is a {\em quantum
correction} to the classical generating function, which is just $%
F_{classical}=(q-Q)^2/2t$. This quantum correction serves to cancel the
second $q$ derivative term in the quantum equation for $F$, ${\frac \partial
{\partial t}}F-{\frac i2}~{\frac{\partial ^2}{\partial q^2}}F+{\frac 12}~{%
\left( {\frac \partial {\partial q}}F\right) }^2=0.$ Immediately, we see the
generalization to any theory for which the propagator is known. The
logarithm of the exact propagator is ($i$ times) an exact time dependent
quantum generating function for the theory. Thus we may obtain $F$ for a
one-dimensional system defined by any quadratic Hamiltonian, or for the
Coulomb problem, and transform the dynamics into $H_0=0$.

However, for the Liouville theory we do not have the exact propagator.
Rather, we are given (\ref{CurvedF}) which tells us how to transform
directly from the exponentially interacting theory to a ``free'' theory on a
curved manifold. As a very simple analogy, let us consider a particle in a
linear potential with a time-dependent coefficient,\ $f(t)$. The
time-dependent Hamiltonian is
\begin{equation}
\label{f(t)q}H(t)=\frac 12\,\,p^2+\,\,f(t)\,\,q.
\end{equation}
Time-dependent wave functions for this system can be obtained from
free-particle wave functions governed by the Hamiltonian
\begin{equation}
\label{Psquare}H_{free}=\frac 12\,P^2.
\end{equation}
A time-dependent generating function $F_2(q,P;t)$ which provides the
connection between the two Hamiltonians is
\begin{equation}
\label{F2}F_2(q,P;t)=\left( q-q_c(t)\right) \left( P+p_c(t)\right)
+S_c(t)\;,
\end{equation}
where $q_c(t)$ and $p_c(t)$ represent {\em any particular classical solution}
to the ``specified acceleration'' problem,
\begin{equation}
dq_c/dt=p_c,~~~~~dp_c/dt=-f(t),
\end{equation}
and where $S_c$ is the corresponding classical action.
\begin{equation}
\label{Sc}S_c(t)={\frac 12}~q_c(t)~p_c(t)-{\frac 12}~\int_0^td\tau ~f(\tau
)~q_c(\tau ).
\end{equation}
Again, we check the {\em classical }requirements on $F_2(q,P;t)$ for it to
implement the necessary canonical transformation. Thus
\begin{equation}
p={\frac \partial {\partial q}}~F_2(q,P;t)=P+p_c,~~~~~Q={\frac \partial
{\partial P}}~F_2(q,P;t)=q-q_c,
\end{equation}
and by directly substituting these
\begin{equation}
H(t)={\frac 12\,}P^2+p_cP+fQ+{\frac 12\,}p_c^2+fq_c=H_{free}-{\ \frac
\partial {\partial t}}~F_2\;.
\end{equation}
We may now clarify how one arrives at the above form for $F_2(q,P;t)$. We
simply observe that {\em any solution }of the classical specified
acceleration problem differs from a particular solution only by a free
particle solution. Hence, $P=p-p_c(t)$ and $Q=q-q_c(t)$ are indeed free
particle variables.

We use $F_2$ in the quantum theory as before. It connects a general
time-dependent wave function $\Phi $$(q,t)$ for a particle governed by (\ref
{f(t)q}) to a free particle wave function $\Psi (P,t)$ through the integral
relation
\begin{equation}
\Phi (q,t)=\frac 1{2\pi }\int dP\;e^{iF_2(q,P;t)}\;\Psi (P,t).
\end{equation}
To convert this to the previous type of generating functional, $F(q,Q;t)$,
we carry out a Fourier transform. So
\begin{equation}
e^{iF(q,Q;t)}=\frac 1{2\pi }\int dP\;e^{-iQP}\;e^{iF_2(q,P;t)}=e^{i\left(
q-q_c(t)\right) p_c(t)+iS_c(t)}\;\delta \left( Q-q+q_c(t)\right) .
\end{equation}
The presence of a Dirac delta function in this result is perhaps surprising.
Nonetheless, it is straightforward to check that $e^{iF(q,Q;t)}$ obeys the
time-dependent partial differential equation
\begin{equation}
\label{(H-id/dt)expiF}\left( H(t)-i{\frac \partial {\partial t}}\right)
\;e^{iF(q,Q;t)}=H_{free}\;e^{iF(q,Q;t)}.
\end{equation}
Thus the relation between $\Phi $$(q,t)$ and a free particle wave function $%
\Psi (Q,t)$ is
\begin{equation}
\label{packet}\Phi (q,t)=\int dQ\;e^{iF(q,Q;t)}\;\Psi (Q,t)=e^{i\left(
q-q_c(t)\right) p_c(t)+iS_c(t)}\;\Psi (q-q_c(t),t).
\end{equation}
A straightforward check shows that $\Phi (q,t)$ as given by this expression
does indeed obey the time-dependent Schr\"odinger equation, $\left( -\frac
12\;\frac{\partial ^2}{\partial q^2}+f(t)\,q\right) \Phi (q,t)=i\frac
\partial {\partial t}\Phi (q,t),$ provided that $\Psi (Q,t)$ is a solution
of the free wave equation, $\left( -\frac 12\;\frac{\partial ^2}{\partial q^2%
}\right) \Psi (q,t)=i\frac \partial {\partial t}\Psi (q,t)$.

We may also use $e^{iF}$ to transform the complete set of time-dependent
free particle wave functions $\Psi _k(Q,t)=e^{ikQ-{\frac i2}k^2t}.$ The
result is to obtain a complete set of time-dependent interacting particle
wave functions $\Phi _k(q,t)=e^{iS_c(t)+i(p_c(t)+k)(q-q_c(t))-{\frac i2}%
k^2t} $. Of course, when $f(t)\rightarrow f$, a constant independent of $t$,
it is possible to obtain the usual stationary state wave functions from $%
\Phi _k(q,t)$ by projection $\Phi _k(q,E)=\int dt~e^{iEt}~\Phi _k(q,t).$

The result given by Eqn.(\ref{packet}) is quite amusing, and not readily
found in the literature. It is related to an old result of Schr\"odinger,
however. To see this, it is useful to generalize a bit to include a
quadratic term in the potential, again with a time-dependent coefficient.
\begin{equation}
\label{Hoscillator}H_{oscillator}(t)=\frac 12\,\,p^2+\frac
12\,k(t)\,q^2+\,\,f(t)\,\,q.
\end{equation}
Now, any solution for this shifted oscillator system is related to a
solution of the unshifted (i.e. $f=0$ ) oscillator by the same canonical
transformation generated by (\ref{F2}), only now the particular classical
solution used in $F_2$ must be a solution of the classical equations
\begin{equation}
\label{oscillator}dq_c/dt=p_c,~~~~~dp_c/dt=-k(t)\,q_c-f(t).
\end{equation}
The action has the same form as in (\ref{Sc}). The same relation (\ref
{packet}) between time-dependent wave packets also applies, only now $\Phi $
and $\Psi $ are solutions of
\begin{equation}
\left( -\frac 12\;\frac{\partial ^2}{\partial q^2}+\frac
12\,k(t)\,q^2+f(t)\,q\right) \Phi (q,t)=i\frac \partial {\partial t}\Phi
(q,t)
\end{equation}
and
\begin{equation}
\left( -\frac 12\;\frac{\partial ^2}{\partial q^2}+\frac
12\,k(t)\,q^2\right) \Psi (q,t)=i\frac \partial {\partial t}\Psi (q,t).
\end{equation}
This relation between wave packets, Eqn.(\ref{packet}), is the
generalization to the time-dependent case of the well-known shift of
variable that relates energy eigenstates for the potential $kq^2/2$ to those
for the potential $fq+kq^2/2$, with $k$ and $f$ constant, namely $%
q\rightarrow q-f/k$. Accompanying this shift of variable, there is also a
shift of energy eigenvalue: $E\rightarrow E-f^2/2k$. The corresponding
`energy' effect in the general time-dependent case is the phase that appears
on the RHS of Eqn.(\ref{packet}). At this point perhaps we should clarify an
issue which should have occurred to the reader. What happens if a {\em %
different }classical solution is used to implement the canonical
transformation? Well, clearly, the difference between any two solutions of
Eqn.(\ref{oscillator}) is a solution of ${\frac{d^2q_d(t)}{dt^2}}%
=-k(t)q_d(t) $, the unshifted oscillator equation. Now the reader may easily
check that any solution of the unshifted oscillator Schr\"odinger equation
remains a solution of that equation if its spatial argument is translated by
$q_d(t)$, and it is multiplied by the phase $e^{iS_d(t)+ip_d(t)(q-q_d(t))}$
(cf. Eqn.(\ref{packet}) with $f=0$). This is just a generalization to the
case of an arbitrary time-dependent $k(t)$, and an arbitrary wave-packet, of
the well-known feature for the time development of the ``minimal Gaussian
packet'' of the harmonic oscillator \cite{Schrodinger1926}.

It remains to further develop these simple ideas involving time-dependent
canonical transformations in quantum mechanics and apply them to analyze the
Liouville theory on a curved surface. We hope to pursue these matters
in the future.

\vskip 0.6cm \noindent ACKNOWLEDGEMENTS

We thank Dr. T. McCarty for his collaboration on various portions of this
research. This work was supported in part by the National Science
Foundation, under grant PHY-90 07517.

\newpage\

\end{document}